\documentclass[twocolumn,showpacs,amsmath,aps]{revtex4}
\usepackage{graphicx,color}
\usepackage{CJK}
\usepackage{bm}
\usepackage[hypertex]{hyperref}
\usepackage{easybmat}

\newcommand{\be}{\begin{equation}}
\newcommand{\ee}{\end{equation}}
\newcommand{\bea}{\begin{eqnarray}}
\newcommand{\eea}{\end{eqnarray}}
\newcommand{\bsube}{\begin{subequations}}
\newcommand{\esube}{\end{subequations}}

\newcommand{\Eq}[1]{Eq.\,(\ref{#1})}

\newcommand{\nl}{\nonumber \\}

\newcommand{\beq}{\begin{equation}}
\newcommand{\eeq}{\end{equation}}
\newcommand{\beqn}{\begin{eqnarray}}
\newcommand{\eeqn}{\end{eqnarray}}
\newcommand{\bsub}{\begin{subequations}}
\newcommand{\esub}{\end{subequations}}

\begin{document}
\begin{CJK*}{GBK}{Song}

\title{The optimal shape of an object for generating
        maximum gravity field \\
        at a given point in space}

\author{Xiao-Wei Wang}
\affiliation{Department of Physics, Beijing Normal University,
Beijing 100875, China}
\author{Yue Su}
\affiliation{Department of Physics, Beijing Normal University,
Beijing 100875, China}

\begin{abstract}
How can we design the shape of an object, in the framework of Newtonian gravity,  in order to generate maximum gravity at a given point in space?
In this work we present a study on this interesting problem.
We obtain compact solutions for all dimensional cases.
The results are commonly characterized by a simple ``physical" feature
that any mass element unit on the object surface generates
the {\it same} gravity strength at the considered point,
in the direction along the rotational symmetry axis.
\end{abstract}


\maketitle

\date{\today}


{\flushleft Newton's gravity} resembles the Coulomb force in that
the strengths of both forces obey an inverse square law \cite{Newton}.
However, unlike the electric charges which can be both
positive and negative, we have only positive mass in our Universe.
Thus there is no gravitational screening effect.
For a metallic body with free charges,
we know that the strongest electric field
is at the place with maximum curvature \cite{Jackson}.
Obviously there is no such a phenomenon for gravity.
However, we may ask a more interesting question:
what is the optimal shape of an object, 
i.e., by properly designing the mass distribution 
under the constraints of fixed total mass and (homogeneous) mass density, 
such that it generates maximum gravity at a given point in space?

For three-dimensional Euclidean space, one might simply guess that
it is a {\it sphere}, as in the case of dewdrops for which
a spherical shape is optimal, which is determined 
by minimizing the surface area (or equivalently, surface energy) 
under the constraint of fixed volume (the same constraint as 
our present one).
However, for the ``maximum-gravity" problem, 
the maximally symmetric shape turns out not to be the optimal!
We may argue this by the following simple consideration.
For a sphere with mass density $\rho$ and radius $R$, the gravity
strength at any point on the sphere is $g=(4\pi/3)\rho GR$.
Let us choose without loss of generality a point ``$A$" on the surface
(see Fig.\ 1) and then remove out a {\it small mass ball}
(with radius $\epsilon\rightarrow 0$) on the opposite side near point ``$B$".
Then, the gravity strength at ``$A$" generated by the remaining matter is
\bea
g^{(1)} &=& \frac{4\pi}{3}\rho G R - \frac{4\pi}{3}\rho G
\frac{\epsilon^3}{(2R-\epsilon)^2}  \nl
&\simeq& \frac{4\pi}{3}\rho G \left(R-\frac{\epsilon^3}{4R^2} \right) \,,
\eea
where we have expanded the result to the next subleading order in $\epsilon$.
On the other hand, if we deform this reduced object to be a new sphere,
the gravity strength on the surface will be
\bea
g^{(2)} \simeq \frac{4\pi}{3}\rho G \left(R-\frac{\epsilon^3}{3R^2} \right) \,.
\eea
It is now clear that for sufficient small $\epsilon$, we have $g^{(1)}>g^{(2)}$.
The naive expectation that a sphere is optimal is wrong.

\begin{figure}
  \centering
  \includegraphics[scale=0.4]{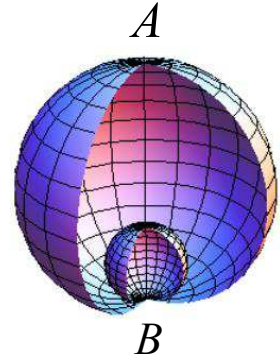}
  \caption{(color online)
  Schematic plot for an argument that a spherical shape is not optimal
  for generating maximum gravity on the surface (e.g., at the point ``$A$").  }
\end{figure}

We are therefore motivated in this work to search for the {\it optimal}
shape of the object, first for the three-dimensional (3-D)
case (that is relevant in our Universe), then for two-dimensional (2-D) case
(in the Appendix we also consider the arbitrary $n$-dimensions
as a math generalization).
As we will see, the 2-D solution is indeed a {\it circular disk} that fits the naive expectation. We find that in all dimensions the solutions are
characterized by the simple physical picture that the optimal shape is axial symmetric, and any matter element unit on the surface
contributes equally to the gravity field
at the considered point in the direction along the axial symmetric axis.

\vspace{0.2cm}
{\flushleft {\bf Symmetry Considerations.}--- }
To get insight for the optimal shape (mass distribution), 
we first present a qualitative {\it symmetry} analysis. 
Let us imagine to fill the matter into the (2-D) circular 
or (3-D) spherical shells as schematically shown in Fig.\ 2.
First, for each shell, in order to generate maximal gravity 
at the point ``$O$" (the origin of the coordinate system), 
the matter should be filled as close as
as possible (however, under the condition of fixed mass density).
This means that there should be no empty space (e.g., holes)
in the mass manifold.
Second, consider the gravity contributions at ``$O$"
from the total mass in all the shells. 
Note that the gravity at ``$O$" generated by the matter 
in each shell is along the direction from ``$O$" to the mass center.
Then, in order to obtain maximum gravity at ``$O$", 
all these gravity ``vectors" should be pointed 
to the same direction (according to the vector summation rule).
Finally, in each shell, 
we should not fill matter to more than a half volume of the shell,
since this will ``waste" mass to cancel some amount of the gravity 
generated by the mass in the {\it half volume} of the shell.

Therefore, based on these symmetry considerations, 
we conclude that the mass distribution should be
on the same side of the $y$-axis and
must be {\it axisymmetric} with respect to the $x$-axis
in order to generate maximum gravity at ``$O$".

\begin{figure}
  \centering
  \includegraphics[scale=0.55]{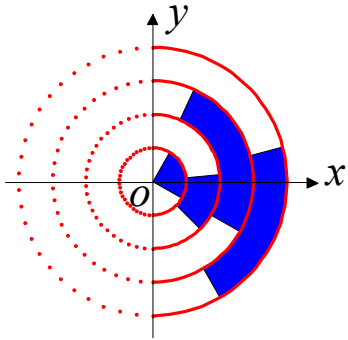}
  \caption{(color online)
Schematic plot for {\it symmetry analysis} 
for mass distribution of an object 
in order to generate maximum gravity at ``$O$".
The plot guides us to imagine to fill the matter (blue thing) 
into the circular or spherical shells. 
Careful symmetry consideration indicates 
that the mass distribution should be 
on the same side of the $y$-axis and 
must be {\it axisymmetric} with respect to the $x$-axis. }
\end{figure}

\begin{figure}
  \centering
  \includegraphics[scale=0.45]{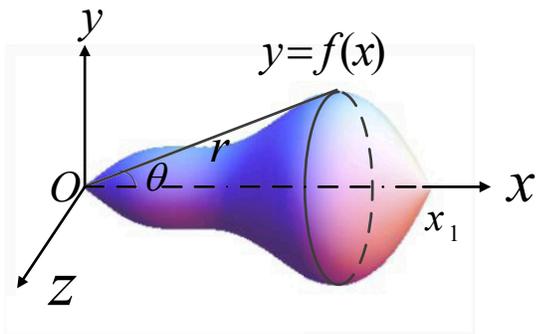}
  \caption{(color online)
  Search for the optimal solution by finding $y=f(x)$
  and rotating it around the $x$-axis, in order to generate maximum gravity
  at the origin ``$O$".  }
\end{figure}

\vspace{0.2cm}
{\flushleft {\bf Three-Dimensional Space.}--- }
Let us consider first the 3-D Euclidean space that is relevant in our Universe.
As shown schematically in Fig.\ 3, an element of mass $dM=2\pi\rho y dx dy$ generates gravity field at the point ``$O$" is
$dg=[G(dM)/r^2]\cos\theta$. From the symmetry consideration, we have projected
the gravity field along the $x$ direction in terms of the factor $\cos\theta=x/r$, where $r=\sqrt{x^2+y^2}$.
Then, the total gravity strength at point ``$O$" is given by
\bea
g = 2\pi\rho G \int^{x_1}_{0}\int^{y}_{0} \frac{xy}{r^3} dx dy  \,.
\eea
More explicitly, we have
\bea
g = 2\pi\rho G \int^{x_1}_{0} dx (1-x/r)
\equiv \int^{x_1}_{0} dx F[x,y(x)]  \,.\label{generalg}
\eea
The total mass of the object can also be calculated, giving
\bea\label{mass-1}
M = \pi\rho \int^{x_1}_{0} dx y^2
\equiv \int^{x_1}_{0} dx \Phi[x,y(x)] \,.
\eea
In our setup, the total mass is a fixed number.  Thus our problem reduces to find the function $f(x)$ that generates the maximum $g$ in (\ref{generalg}) while keeping the mass (\ref{mass-1}) fixed.

Based on the variational principle \cite{math},
we adopt the Lagrange multiplier method
and define a new functional:
\be
H = F + \lambda \Phi,
\ee
where $\lambda$ is the Lagrangian multiplier.  
Note that for variational problems with constraints, 
after introducing the Lagrangian multiplier $\lambda$, 
one can safely obtain the Euler-Langrange equation, 
by performing the variational calculus.
Also, we may point out that the present problem differs from 
the usual variational problem with fixed-end-points, 
since the constraint of fixed total mass or volume implies 
the movable-end-point at $x_1$.
However, this does not affect the use of the variational principle 
which allows us to obtain the Euler-Lagrange equation,
$H_y-\frac{d}{dx}(H_{y'})=0$, yielding
\bea\label{res-1}
\frac{x}{(x^2+y^2)^{3/2}}=-\frac{\lambda}{G} \,.
\eea
This equation determines the shape of the object, namely
$y=f(x)$, as shown in Fig.\ 3.

The parameter $\lambda$ can be determined by the constraint equation
\Eq{mass-1}. Substituting the shape function \Eq{res-1}
into \Eq{mass-1}, together with the boundary
condition $y(x_1)=0$ which gives $x_1=(-G/\lambda)^{1/2}$,  we find
\bea
\lambda = -G \left( \frac{4\pi\rho}{15M} \right)^{2/3} \,.
\eea
Accordingly, the maximum gravity (strength) at point ``$O$" can be integrated, giving
\bea\label{gmax}
g_{\rm max} = \frac{4}{5}\pi\rho G \left(\frac{15M}{4\pi\rho} \right)^{1/3} \,.
\eea
It is instructive to compare this result with some other shapes. 
In Fig.\ 4, we compare it with two examples of shapes that admit
local maxima.  In Fig.\ 5, we consider deforming the object by introducing a parameter $a$ so that the shape is now described by $x(ax^2+y^2)^{-3/2}=-\lambda/G$. 
The shape deviates from the optimal solution \Eq{res-1} via $a\neq 1$.
In both plots we see that 
$g_{\rm max}$ is indeed the maximum. 
One may also compare $g_{\rm max}$ with the result generated by a sphere,
$(4\pi/3)^{2/3}\rho G (M/\rho)^{1/3}$,
thus finding that $g_{\rm max}$ is greater by about 2.6\%.

While these numerical comparisons might be incomplete, 
we would like to remark that \Eq{res-1}
is the {\it overall} optimal solution,
which is guaranteed by both the {\it symmetry} 
and {\it variational} principles. 
The above ``incomplete" comparisons are just for intuitive illustration.

\begin{figure}
  \centering
  \includegraphics[scale=0.5]{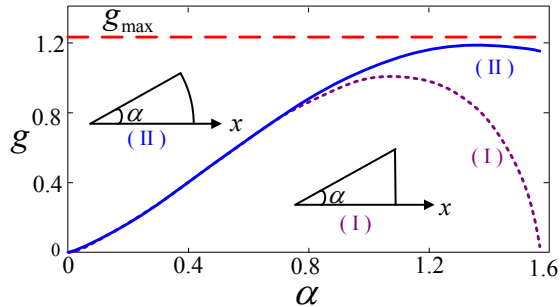}
  \caption{(color online)
  Comparison of the maximum gravity $g_{\rm max}$, given by \Eq{gmax},
  with results from two other restricted shapes as shown in the inset,
  by rotating a ``triangle" (I) and a ``sector" (II) around the $x$-axis.
  For these two shapes, under the mass constraint condition,
  we change the angle $\alpha$ to show the resulted gravity
  at the origin. Indeed, we see that $g_{\rm max}$  is the maximum.
  In the plots, $g$ has been scaled by $G(\pi^2\rho^2M)^{1/3}$.     }
\end{figure}

\begin{figure}
  \centering
  \includegraphics[scale=0.58]{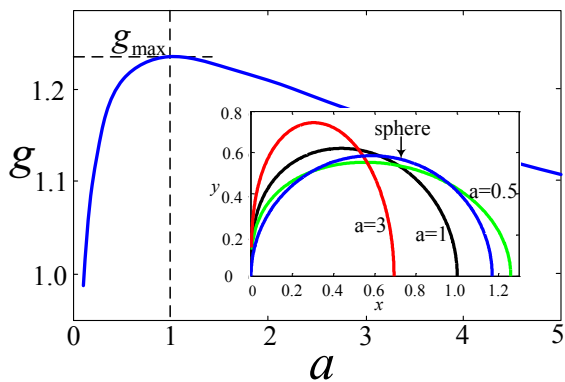}
  \caption{(color online)
  Plot for showing the optimal condition by deforming
  the object as $x(ax^2+y^2)^{-3/2}=-\lambda/G$,
  which deviates from the optimal solution \Eq{res-1}
  via $a\neq 1$. Similar as in Fig.\ 4,
  $g$ has been scaled by $G(\pi^2\rho^2M)^{1/3}$.
Inset: comparison of the {\it optimal} shape given by Eq.\ (10)
(the black curve with $a=1$)
with two less optimal shapes ($a\neq 1$)
and also with the sphere.  }
\end{figure}

To get further insight, we re-express the 
shape function \Eq{res-1} as
\bea
\cos\theta / R^2 = {\rm const} \,,
\eea
where $R$ is the distance from the origin ``$O$" to the
{\it surface} point of the object at the angle $\theta$.
We may characterize this elegant result in more 
``physical" conceptual way: any mass element unit on the object surface
contributes equally to the gravity field at the origin ``$O$"
along the rotational symmetry axis.

\vspace{0.2cm}
{\flushleft {\bf Two-Dimensional Space.}---}
As an interesting comparison, 
we consider now the 2-D space.
The physical realization for the 2-D case 
is an infinite cylinder and the 2-D mass element  
corresponds to an infinitely long wire, 
which generates the gravity strength of $\sim 1/r$
at the distance $r$ (away from the wire).  
Using also Fig.\ 3 but without rotating around the $x$ axis,
we express the gravity strength at the origin
and the mass constraint condition as:
\bea
g &=& \int^{x_1}_{0}\int^{y}_{0} \, G\rho \frac{\cos\theta}{r} dx dy
 = G\rho \int^{x_1}_{0} \arctan\frac{y}{x} dx   \nl
 &\equiv&  \int^{x_1}_{0} F[x,y(x)] dx  \,,   \\
\nl
M &=& \int^{x_1}_{0} \rho y dx
\equiv \int^{x_1}_{0} \Phi[x,y(x)] dx    \,.
\eea
Applying the Lagrange multiplier method \cite{math},
we introduce the auxiliary functional $H=F+\lambda \Phi$.
Then, with respect to $H[x,y(x)]$, the Euler-Lagrange equation gives
\bea
G \frac{x}{x^2+y^2} + \lambda = 0 \,,
\eea
or in a more compact form,
\bea\label{res-2d}
\frac{\cos\theta}{R} = - \frac{\lambda}{G} \,.
\eea
Here we use $R$ to denote the point on the surface of the object.
The Lagrange multiplier $\lambda$ can be easily determined
from the mass constraint, as done in the 3-D case.

\Eq{res-2d} tells us that a {\it circular disk} is the optimal shape
of the object in 2-D space under the constrained condition. This result happens to agree with the naive expectation that the most symmetric shape is optimal. However, the 3-D example demonstrated that this does not generalize to other dimensions.\\
\\
To summarize, we have presented an intriguing study on 
a gravity extremum problem. 
By careful symmetry analysis, we first determine that 
the optimal shape must be {\it axisymmetric}. 
We then employ the variational method to search for 
the optimal mass distributions for 3-D and 2-D objects, 
respectively, while for the arbitrary $n$-dimensions in Appendix.
We obtain compact solutions for all dimensional cases
and find that all the solutions are
characterized by the simple picture that 
any matter element unit on the surface
contributes equally to the gravity field
at the considered point in the direction along the axial symmetric axis.

\vspace{0.6cm}
{\it Acknowledgments.}---
The authors are grateful to Hong L\"u for introducing
the problem, to Xin-Qi Li for discussions
and to Luting Xu for technical assistance.
This work was supported by the Undergraduate Student
Training Project of Beijing Normal University.

\appendix

\section{Arbitrary $n$-Dimensional Space}

In this Appendix we generalize our study in the main text 
to arbitrary $n$-dimensions. 
For intuitive purpose, we refer again to Fig.\ 3.
In the $n$-dimensional space, the rotating ``circle"
in Fig.\ 2 corresponds to an $(n-2)$-dimensional sphere,
thus the mass element reads $dM=\rho (\Omega y^{n-2})dxdy$.
Here we have denoted
the volume factor of the $(n-2)$-dimensional sphere
by $\Omega\equiv \pi^{n/2-1}/\Gamma(\frac{n}{2})$,
where $\Gamma(\bullet)$ is the Gamma function.
Also, in the $n$-dimensional space, the gravity strength
is proportional to $1/r^{n-1}$.
Then, from the gravity strength at the origin
$g = \int^{x_1}_{0}\int^{y}_{0}dM \; G\cos\theta/r^{n-1}$,
and the mass constraint
$ M = \int^{x_1}_{0}\int^{y}_{0} dM $,
we extract two functionals as
\bea
F[x,y(x)] &=& G\rho \Omega \int^{y}_{0} dy \, x y^{n-2}/r^n  \,,  \\
\Phi[x,y(x)] &=& \rho \Omega \int^{y}_{0} dy \, y^{n-2} \,.
\eea

As in the earlier 3-D and 2-D examples, we introduce $H=F+\lambda \Phi$
and apply the Euler-Lagrange equation \cite{math}. We find
\bea\label{res-n1}
G\left( \frac{xy^{n-2}}{R^n}  \right)
+ \lambda  \, y^{n-2} = 0  \,,
\eea
where, as in the above, $R$ (not $r$)
denotes the distance of the surface point to the origin.
From this equation we obtain
\bea\label{res-n2}
\frac{x}{R^n}=-\frac{\lambda}{G} ~~~ {\rm or} ~~~
\frac{\cos\theta}{R^{n-1}}=-\frac{\lambda}{G} \,.
\eea
Inserting this result into the mass constraint condition,
the multiplier $\lambda$ can be determined.

The elegant formula \Eq{res-n2} generalizes our previous 
three- and two-dimensional results 
presented in the main text to arbitrary dimensions.
As we find, it continues to hold the simple physical picture:
any mass element unit on the shape surface contributes equally
to the gravity field at the origin,
projected onto the direction of the axial symmetric axis.

\end{CJK*}
\end{document}


\begin{CJK*}{GBK}{Song}

\title{ A supplemental note to ``the optimal shape
of an object for generating \\
maximum  gravity field
at a given point in space" }

\author{Xiao-Wei Wang}
\affiliation{Department of Physics, Beijing Normal University,
Beijing 100875, China}
\author{Yue Su}
\affiliation{Department of Physics, Beijing Normal University,
Beijing 100875, China}

\begin{abstract}
In this note we provide a simple analysis
to conceptually understand the main result
derived in our recent work
[Eur. J. Phys. {\bf 36}, 055010 (2015)]
on the ``gravity extremum" problem.
\end{abstract}


\maketitle
\date{\today}

{\flushleft In our recent work \cite{Wang15}, }
we presented a study on
the interesting problem how to optimally design
the shape of an object, in the framework of Newtonian gravity,
in order to generate maximum gravity at a given point in space.
Using the variational method and involving no assumptions,
we obtained an elegant solution to the problem
for all dimensional cases.  \\
\\
The problem addressed in Ref.\ \cite{Wang15} is of particular
interest to undergraduate students and useful for education.
Very recently, this problem was revisited
by Datsenko, Lozovenko and Minaiev (DLM) \cite{DLM16}
and was connected with the Little Prince's problem
(see The Little Prince by Antoine de Saint-Exup\'ery):
``{\it the prince wanted to change the shape of his
asteroid for generating maximum gravity field
at a point where his rose was growing}".
In Ref.\ \cite{DLM16},
the {\it optimal shape equation} (OSE)
or {\it equal gravity condition} (EGC)
we derived in Ref.\ \cite{Wang15}
was used as an assumption or starting point,
as stated by the authors:
`` {\it Consider the shape is already known and looks like the
one shown in figure 1 (our students suggested this form) ...
If this statement is true, then projections of gravitational
strength on the axis $z$ generated
by the point mass $m$ at the point
$A$ are equal for any position of the point mass (see figure 1)}".
Then, they solve the problem using only simple calculus
for the three-dimensional space. \\
\\
Noting that the OSE or EGC is the central result of Ref.\ \cite{Wang15}
and the starting point of Ref.\ \cite{DLM16}, in this supplement
we provide a simple understanding (proving) to the EGC.
For simplicity, we use the short term EGC
to denote the condition of generating
{\it equal $z$-axial component of gravity}
at the point ``$O$" by any mass element $dm$
on the planet surface,
where $z$ is the rotational axis and ``$O$"
the point under consideration
(for maximum gravity field generated).
Obviously,
{\it the optimal shape should satisfy
the sufficient-and-necessary condition
that any move of mass (change of shape)
must not increase the gravity at the point ``$O$" }.  \\
\\
For the sake of simplicity, let us consider only
moving a small mass element ($dm$).
From the {\it incompressible} condition, we know that the mass moving
can be performed only from the inside of the planet to its outside.
First, let us consider a planet which satisfies the EGC.
{\it (i)}
If we move a mass $dm$, for instance,
from a surface point ``$A_1$" to another (arbitrary) surface point ``$A_2$",
the EGC tells us that the gravity at the point ``$O$" does not change.
{\it (ii)}
For a bit more complicated case,
consider an inside point ``$A$",
which has the same distance to the point ``$O$"
as the surface point ``$A_1$" has
(note that there are many--infinite number--of such points).
If we move a mass $dm$ from the inside point ``$A$"
to the (arbitrary) surface point ``$A_2$",
the $z$-axial component of gravity at ``$O$"
(generated by the $dm$) will become smaller,
since the gravity field at ``$O$" generated by
the mass $dm$ at point ``$A$"
has a larger $z$-axial component than that
generated by the same mass at the surface point ``$A_1$".  \\
\\
Second, alternatively, let us consider
certain possible solution (if it exists) not satisfying the EGC.
In this case, on the planet surface,
there must exist ``better" and ``worse"
places where the same $dm$ does not generate
equal $z$-axial component of gravity at the point ``$O$".
To be specific, let us denote the ``better" place by ``$A_2$",
and the ``worse" place by ``$A_1$".
Similarly as done in the above {\it (ii)},
associated with the surface point ``$A_1$",
we choose an inside point ``$A$" (from the many ``equal distance" points).
Then, we easily understand that {\it (i)}
moving $dm$ from the surface point ``$A_1$" to ``$A_2$"
will enhance the gravity at ``$O$",
while {\it (ii)} the mass move from the inside point ``$A$"
to the surface point ``$A_2$"
will make the result uncertain
(can enhance or reduce the gravity at the point ``$O$").  \\
\\
Therefore, through the above analysis, we actually proved
that the EGC is the {\it sufficient-and-necessary} condition.
The above consideration is based on moving a small mass $dm$.
One may question that
the resultant change of gravity at ``$O$" is negligibly small.
Indeed, more mass (thus larger volume) moving will make
the analysis more complicated.
Nevertheless, the above qualitative analysis is a valuable
addition to our previous mathematical treatment \cite{Wang15}.
With the EGC as a starting point, as done in Ref.\ \cite{DLM16},
one can straightforwardly perform the remaining simple calculus
to get the full results, without further intelligence. \\
\\
\\
{\bf Acknowledgments}.---
The authors are grateful to Xin-Qi Li
and Hong L\"u for suggestion and discussion.

\end{CJK*}